\documentclass[preprint,aps]{revtex4}

\usepackage{graphicx}

\begin{document}

\title{Direct Evidence of Interaction-Induced Dirac Cones in Monolayer Silicene/Ag(111) System}

\author{Ya Feng$^{1,\sharp}$, Defa Liu$^{1,\sharp}$, Baojie Feng$^{1,\sharp}$, Xu Liu$^{1,\sharp}$, Lin Zhao$^{1}$, Zhuojin Xie$^{1}$, Yan Liu$^{1}$, Aiji Liang$^{1}$, Cheng Hu$^{1}$, Yong Hu$^{1}$, Shaolong He$^{1}$, Guodong  Liu$^{1}$, Jun Zhang$^{1}$, Chuangtian Chen$^{2}$, Zuyan Xu$^{2}$,  Lan Chen$^{1}$,  Kehui Wu$^{1,3}$, Yu-Tzu Liu$^{4,5}$,  Hsin Lin$^{4,5}$, Zhi-Quan Huang$^{6}$, Chia-Hsiu Hsu$^{6}$, Feng-Chuan Chuang$^{6}$, Arun Bansil$^{7}$ and X. J. Zhou$^{1,3,*}$
}

\affiliation{
\\$^{1}$Beijing National Laboratory for Condensed Matter Physics, Institute of Physics, Chinese Academy of Sciences, Beijing 100190, China
\\$^{2}$Technical Institute of Physics and Chemistry, Chinese Academy of Sciences, Beijing 100190, China
\\$^{3}$Collaborative Innovation Center of Quantum Matter, Beijing 100871, China
\\$^{4}$Centre for Advanced 2D Materials and Graphene Research Centre, National University of Singapore, Singapore 117546
\\$^{5}$Department of Physics, National University of Singapore, Singapore 117542
\\$^{6}$Department of Physics, National Sun Yat-Sen University, Kaohsiung 804, Taiwan
\\$^{7}$Department of Physics, Northeastern University, Boston, Massachusetts 02115, USA
}

\date{March 21, 2015}

\pacs{}

\maketitle

{\bf Silicene, analogous to graphene, is a one-atom-thick two-dimensional crystal of silicon which is expected to share many of the remarkable properties of graphene.  The buckled honeycomb structure of silicene, along with its enhanced spin-orbit coupling, endows silicene with considerable advantages over graphene in that the spin-split states in silicene are tunable with external fields. Although the low-energy Dirac cone states lie at the heart of all novel quantum phenomena in a pristine sheet of silicene, the question of whether or not these key states can survive when silicene is grown or supported on a substrate remains hotly debated.  Here we report our direct observation of Dirac cones in monolayer silicene grown on a Ag(111) substrate.  By performing angle-resolved photoemission measurements on silicene(3$\times$3)/Ag(111), we reveal the presence of six pairs of Dirac cones on the edges of the first Brillouin zone of Ag(111), other than expected six Dirac cones at the K points of the primary silicene(1$\times$1) Brillouin zone. Our result shows clearly that the unusual Dirac cone structure originates not from the pristine silicene alone but from the combined effect of silicene(3$\times$3) and the Ag(111) substrate. This study identifies the first case of a new type of Dirac Fermion generated through the interaction of two different constituents. Our observation of Dirac cones in silicene/Ag(111) opens a new materials platform for investigating unusual quantum phenomena and novel applications based on two-dimensional silicon systems.}

Silicene is theoretically predicted to be stable in the honeycomb lattice and, similar to graphene, it exhibits the characteristic low-energy Dirac cone state\cite{KTakeda_prb,GGGuzma_prb,SCa_prl,SLebegue,Voon}. Silicene can thus be expected to share most of the remarkable quantum properties of graphene\cite{gra_sci,QHE1_Nat,QHE2_Nat,Klein_RMPhys}. Distinct from graphene, however, which is essentially flat dominated by sp$^{2}$ bonding, the crystal structure of silicene is buckled with mixed sp$^{2}$/sp$^{3}$ bonding\cite{KTakeda_prb,GGGuzma_prb,SCa_prl,SLebegue}. The much stronger spin-orbit coupling in silicene\cite{CCLiu_prl,CCLiu_prb} leads to a larger energy gap at the Dirac points and makes it possible to realize the quantum spin Hall effect in an experimentally accessible temperature regime\cite{CCLiu_prl,CCLiu_prb}. The buckled honeycomb structure drives a number of new phenomena and properties in silicene. In particular, the gap at the Dirac point can be tuned by applying external electric and magnetic fields to realize a variety of different phases and topological phase transitions\cite{ZYN_Tgap,NDDrummond_PRB,MEzawa_NJP,MEzawa_prl,LStille_Tgap}. The unique advantages of silicene and its compatibility with the traditional silicon industry make it an attractive materials platform for next generation nanoelectronics applications\cite{WFT_ncm,XTAn_NJPhy,FLiu,nanodisk,supcurrent,AKara}.

Single-layer and multilayer silicenes have been grown on various supporting materials, with the Ag(111) surface being the most common substrate\cite{BLalmi,Vogt_prl,CLLin,diboride,BJFeng,H.Jam,LMengIr,HJamgotchian}. The buckled structure of silicene naturally leads to the formation of a variety of configurations beyond the primary (1$\times$1) structure under different preparation conditions, such as the (3$\times$3)/Ag(111) and ($\sqrt3\times\sqrt3$)R30$^{\circ}$/Ag(111) structures\cite{BLalmi,Vogt_prl,CLLin,diboride,BJFeng,H.Jam,LMengIr,HJamgotchian}. Although experimental signatures of Dirac fermions have been reported in silicene, these results are highly controversial and inconclusive\cite{Vogt_prl,LChen_prl,CLLin_noLD,PadovaAPL,JAvila,Padova_multilayer,Feng_acsnano,DTsoutsou_APL,SKMahatha_PRB}. Extensive theoretical work indicates that the interaction between silicene and the Ag(111) substrate will destroy the Dirac cones in silicene\cite{CLLin_noLD,ZXGuo,SCahangirov,YPWang,RQuhe,MXChen}. Here we report our direct observation of Dirac cones in monolayer silicene grown on a Ag(111) substrate.  By performing in-depth angle-resolved photoemission measurements on silicene(3$\times$3)/Ag(111), we reveal the presence of six pairs of Dirac cones on the edges of the first Brillouin zone of Ag(111), not on the K points of the primary silicene(1$\times$1) Brillouin zone. This result shows clearly that the observed Dirac cones originate not from the pristine silicene film alone but from the combined system of silicene(3$\times$3) and the Ag(111) substrate. Our study thus identifies the first case of a new type of Dirac Fermion generated through the interaction of two different constituents. Our demonstration that silicene(3$\times$3)/Ag(111) can harbor Dirac cones provide a new pathway for exploiting two-dimensional silicon system as material platforms for investigating quantum phenomena and potential applications.

The silicene/Ag(111) sample was grown {\it in situ} in an ultra-high vacuum chamber connected directly with an angle-resolved photoemission (ARPES) system. The Ag(111) single crystal was first cleaned by many cycles of Argon ion sputtering and annealing at $\sim$800 K.  Quality of the Ag(111) surface was checked by low energy electron diffraction (LEED) and ARPES measurements on the surface state around the $\Gamma$ point. Band structure of the Ag(111) surface was measured by ARPES for later comparison with supported silicene surface.  Silicene was grown by heating a piece of silicon wafer to directly deposit Si atoms on the pre-heated clean Ag(111) surface following the same procedure as described in Ref. \onlinecite{BJFeng}. The single-layer silicene(3$\times$3)/Ag(111) sample was prepared at the Ag(111) substrate temperature of  470 K.  Monolayer silicene(3$\times$3)/Ag(111) was found to cover most of the surface area while other minor phases can be neglected as determined from the LEED patterns.  Moreover, such a silicene(3$\times$3) structure can only exist as the first layer on Ag(111); subsequent layers result in the formation of the ($\sqrt3\times\sqrt3$)R30$^{\circ}$ phase (see Fig. S1 in Supplementary Materials). We have repeated the growth process, followed by characterization of the film by LEED and scanning tunneling microscope (STM), many times to make sure that the grown sample has a single-layer silicene(3$\times$3) structure. The ARPES results on silicene(3$\times$3)/Ag(111) presented in this study are highly reproducible.  ARPES measurements were carried out in our lab system with a Scienta R4000 electron energy analyzer and a helium discharge lamp which provides a photon energy of 21.218 eV\cite{GDLiu}. The base pressure of the ARPES system is better than 5$\times$10$^{-11}$ Torr. The angular resolution is $\sim$0.3 degree and the energy resolution was set at 20 meV for increasing the measurement efficiency. The Fermi edge of a clean polycrystalline gold specimen connected to the sample was taken as the reference Fermi level. The deposition of potassium on the silicene/Ag(111) surface was realized by depositing potassium {\it in situ} for different times while keeping the sample at a low temperature of $\sim$20 K.

Figure 1a shows a typical STM image of the monolayer silicene(3$\times$3)/Ag(111) phase in which the 3$\times$3 superstructure can be seen clearly\cite{BJFeng}. For convenient reference, Fig. 1b shows the first Brillouin zones of the Ag(111) surface and primary silicene(1$\times$1), along with the folded Brillouin zone of silicene(3$\times$3). Interestingly, silicene(3$\times$3) lattice has a good match with the Ag(111) surface because the first Brillouin zone of Ag(111) accommodates precisely 16 folded Brillouin zones of the silicene(3$\times$3) phase. Fig. 1c gives the constant energy contours of the clean Ag(111) surface at different binding energies. The Fermi surface (Fig. 1c1) is seen to consist of a clear electron pocket around the $\Gamma$ point due to the well-known Shockley surface state\cite{FReinert}, and a large hexagonal Fermi surface sheet along the Ag(111) Brillouin zone edge. As the binding energy increases to 300 meV (Fig. 1c2) and 600 meV (Fig. 1c3), the central surface state disappears while the bulk hexagonal contour keeps its basic shape but shows a slight decrease in area.

The Fermi surface of silicene(3$\times$3)/Ag(111) (Fig. 1d) shows interesting features that clearly set it apart from that of Ag(111) (Fig. 1c). With monolayer silicene coverage, the signal of the Ag(111) surface state pocket around $\Gamma$ completely disappears and that of the Ag(111) bulk states is strongly suppressed although the residual signal is still discernable. These results are consistent with the single-layer growth mode of silicene(3$\times$3) that can fully cover the Ag(111) surface. The Fermi surface topology of silicene(3$\times$3)/Ag(111) exhibits twelve spot structures along the six edges of the Ag(111) first Brillouin zone (Fig. 1d1). With increasing binding energy, these spots grow into approximately triangle-shaped pockets (Fig. 1d2 and 1d3).  At high binding energies, the two pockets on the Ag(111) Brillouin zone edge touch and merge with each other (Fig. 1d3). Direct comparison of Fig. 1d with that for the Ag(111) surface (Fig. 1c), as well as the comparison of the measured band structures (see Fig. S2 in Supplementary Materials), indicates unambiguously that the observed Fermi pockets along the Ag(111) Brillouin zone edges must be associated with the silicene(3$\times$3) structure grown on the Ag(111) surface. As we will show below (Figs. 2, 3 and 4), each strong spot here represents a Dirac cone structure in silicene(3$\times$3)/Ag(111).

Figure 2 clarifies details of how the observed Dirac cones evolve with increasing binding energy. Two independent high-resolution ARPES measurements were carried out to simultaneously cover the two Dirac cones around the M point (Fig. 2a) and the K point (Fig. 2b). Similar to Fig. 1d, with increasing binding energy, the constant energy contours for the two Dirac cones grow from spots at the Fermi level to triangle-shaped pockets with increasing area at higher binding energy. At the binding energy of $\sim$0.4 eV, the two Dirac cones touch each other near the M point and start to merge at higher binding energies.  On the other hand, the  two Dirac cones around the K point (Fig. 2b) experience a similar trend of increasing area with increasing binding energy. However, these two Dirac cones do not touch each other even at 0.6 eV binding energy. The constant energy contour lines at different binding energies are quantitatively shown in Fig. 1c for one Dirac cone; the results for the four Dirac cones in Fig. 2a and 2b are consistent. These results indicate that the twelve Dirac cones observed around the Ag(111) Brillouin zone edges can be divided into six pairs, each of which is centered at the M point of the Ag(111) Brillouin zone edge, as shown schematically in Fig. 2d.

Figure 3 shows the detailed band structure of silicene(3$\times$3)/Ag(111) measured along different momentum cuts. For all the momentum cuts across one Dirac cone (cuts A, B, C and D), one can see two nearly linear bands extending over an energy range of 1 eV, consistent with our picture of the Dirac cones.  For cut A, the two pairs of Dirac bands do not cross up to 1.4 eV, while for cut B along one Ag(111) Brillouin zone edge, the two pairs of Dirac cones intersect at $\sim$0.4 eV binding energy, consistent with the six-pair picture of Fig. 2d. In particular, Fig. 3b indicates that the observed signal is not simply an addition of two individual Dirac cones, ruling out the possibility that one pair of Dirac cones on an edge comes from two domains with different orientations. This is consistent with our LEED results where there is only a 3$\times$3 superstructure along a single orientation (see Fig. S1 in Supplementary Materials). For a given Dirac cone, different cuts give similar linear bands but with different dispersions (Fig. 3c and 3d). Fig. 3g plots the Fermi velocity of one Dirac cone along different orientations; the values are obtained by fitting the bands near the Fermi level along different momentum cuts (Fig. 2(a-c) and Fig. 3(a-d))). It is clear that the Dirac cone in silicene(3$\times$3)/Ag(111) is actually not cone-like in that its cross section at various binding energies is more like a triangle than a circle (Fig. 2 and Fig. 3g). The resulting Fermi velocity is quite anisotropic with an approximate three-fold symmetry, varying from 2 to 4 eV$\cdot$$\AA$ (corresponding to 3 to 6$\times$10$^{5}$m/s)(Fig. 3g). Notably, no Fermi crossing is seen on the measured band in cut E, consistent with the Dirac cone picture of Fig. 2d. The observed bands in cuts B and E are consistent with previous ARPES measurements on silicene(3$\times$3)/Ag(111)\cite{DTsoutsou_APL, SKMahatha_PRB} although Dirac cones were not identified in these earlier studies. We emphasize that greatly suppressed signal of pure Ag(111) surface in our silicene(3$\times$3)/Ag(111) sample played a crucial role in allowing us to reveal the presence of Dirac cones in our samples.


It is clear that the Dirac points of our silicene(3$\times$3)/Ag(111) sample lie above the Fermi level, and therefore cannot be seen at low temperatures because of the Fermi-Dirac cutoff of the photoemission process.  In order to observe the upper Dirac branch and thus the whole Dirac cone, we have employed two different approaches. The first is to warm up the sample to make use of thermal excitation of electrons above the Fermi level (Fig. 4a-d). Dividing out the corresponding Fermi-Dirac distribution function makes it possible to observe a portion of the band structure above the Fermi level at high temperature. As seen in Fig. 4d, the Dirac bands are stable up to 450 K.  Also the Dirac cone is observable at 450 K which is about 170 meV above the Fermi level (Fig. 4d). These results indicate that our silicene(3$\times$3)/Ag(111) sample is hole-doped. An alternative way to reveal the Dirac point and the upper Dirac branch is to deposit potassium onto the silicene(3$\times$3)/Ag(111) surface which is expected to provide electron-doping. As seen in Fig. 4(e-h), with increasing potassium deposition, indeed the Dirac cone shifts downwards as expected. When the potassium doping is high enough (Fig. 4h), the Dirac cone is shifted to nearly 200 meV below the Fermi level and the upper Dirac bands become visible. In this case, the sample has become electron-doped. These results further establish the Dirac cone structure in silicene(3$\times$3)/Ag(111) sample and demonstrate the possibility of transforming the sample from being hole-doped to an electron-doped case.

Our observation of six-pairs of Dirac cones in silicene(3$\times$3)/Ag(111) is unusual for many reasons. Firstly, there can be no doubt that the Ag(111) surface alone does not support such Dirac cones;  these Dirac cones come into existence only after silicene(3$\times$3) is grown on Ag(111). Secondly, the observed six pairs of  Dirac cones are fundamentally different from those for a free-standing honeycomb lattice which would have six Dirac cones at the K points of its own Brillouin zone.  Thirdly, the present six pairs of Dirac cones lie on the edges of the first Brillouin zone of Ag(111) without an obvious connection with the first Brillouin zone of the primary silicene(1$\times$1). Fourthly, the (3$\times$3) superstructure of silicene is expected to induce band-folding and duplicate features in the reduced Brillouin zones (Fig. 1b), but we do not observe indications of such band folding due to the superstructure. The preceding observations imply clearly that the six pair Dirac cone structure we have observed in silicene(3$\times$3)/Ag(111) does not exist in either free-standing silicene(3$\times$3) alone or in the Ag(111) surface alone; it must be an effect generated through the interaction between the pristine silicene film and the Ag(111) substrate when the two systems are combined.  In particular, the low energy electronic states of silicene(3$\times$3)/Ag(111) come from the hybridization of silicene and Ag(111) {\it sp} states with their periodicity being dictated essentially by the Ag(111) lattice.  This is consistent with theoretical predictions\cite{CLLin_noLD,ZXGuo,SCahangirov,YPWang,RQuhe,MXChen} and experimental measurements\cite{DTsoutsou_APL, SKMahatha_PRB} which indicate that Ag(111) substrate interacts strongly with silicene and destroys the Dirac cones at K points of free-standing silicene(1$\times$1)\cite{CLLin_noLD,ZXGuo,SCahangirov,YPWang,RQuhe,MXChen}.  However, even though the silicene/Ag(111) systems have been extensively investigated\cite{CLLin_noLD,ZXGuo,SCahangirov,YPWang,RQuhe,MXChen}, there are no band structure calculations so far, including our own extensive computations (see Fig. S4 in Supplementary Materials), which can explain the unusual six pair Dirac cone structure we have reported here in a silicene(3$\times$3)/Ag(111) sample.

In summary, we have provided direct evidence for the existence of six pairs of Dirac cones in the monolayer silicene(3$\times$3)/Ag(111) system.  We have demonstrated that this unusual Dirac-cone structure only comes into existence when the silicene film is grown on the Ag(111) substrate. Our study thus identifies a new type of Dirac cone structure which is obtained through the interaction of silicene with the substrate to generate a novel state that is distinct from that of their individual constituents. The observed six-pair Dirac cone structure in silicene/Ag(111) system cannot be understood in terms of existing band structure calculations, and we hope that our study will stimulate further related theoretical work. Our observation of a novel Dirac cone structure in silicene(3$\times$3)/Ag(111), and the possibility of hole or electron doping of these Dirac cones, open a new materials pathway for fundamental science investigations and applications based on two-dimensional silicon systems.

$^{\sharp}$These people contributed equally to the present work.

$^{*}$Corresponding authors: XJZhou@aphy.iphy.ac.cn

\vspace{3mm}

\noindent {\bf Acknowledgement} XJZ thanks financial support from the NSFC (91021006, 11334010, 11334011 and 11474336), the MOST of China (973 program No: 2011CB921703, 2011CBA00110, 2012CB821402, 2013CB921700 and 2013CB921904), and the Strategic Priority Research Program (B) of the Chinese Academy of Sciences (Grant No. XDB07020300). The work at Northeastern University was supported by the US Department of Energy (DOE), Office of Science, Basic Energy Sciences grant number DE-FG02-07ER46352 (core research), and benefited from Northeastern University's Advanced Scientific Computation Center (ASCC), the NERSC supercomputing center through DOE grant number DE-AC02-05CH11231, and support (applications to layered materials) from the DOE EFRC: Center for the Computational Design of Functional Layered Materials (CCDM) under DE-SC0012575. H.L. acknowledges the Singapore National Research Foundation for support under NRF Award No. NRF-NRFF2013-03. FCC acknowledges support from the National Center for Theoretical Sciences and the Taiwan Ministry of Science and Technology under Grant Nos. MOST-101-2112-M-110-002-MY3 and MOST-101-2218-E-110-003-MY3, and the support the National Center for High Performance Computing for computer time and facilities.

\vspace{3mm}

\noindent {\bf Author Contributions}\\
Y.F., D.F.L, B.J.F. and X.L. contribute equally to this work.  X.J.Z., Y.F., D.F.L., B.J.F., X.L., L.Z, Z.J.X. and K.H.W. proposed and designed the research. Y.F., D.F.L., B.J.F., X.L., L.Z, Z.J.X. L.C. and K.H.W. contributed in sample preparation. Y.F., D.F.L., X.L., L.Z., Z.J.X., Y.L., A.J.L., C.H., Y.H., S.L.H., G.D.L., J.Z., C.T.C., Z.Y.X. and X.J.Z. contributed to the development and maintenance of Laser-ARPES system. Y.F., D.F.L., X.L., L.Z. and Z.J.X. carried out the ARPES experiment.  Y.F., D.F.L., B.J.F.,  X.L., L.Z, Z.J.X. and X.J.Z. analyzed the data. Y.T.L., H.L., Z.Q.H., C.H.H., F.C.C. and A.B. performed band structure calculations.  X.J.Z., Y.F. and A.B. wrote the paper with D.F.L, B.J.F., X.L. and L.Z., and all authors participated in discussion and comment on the paper.



\newpage

\begin{figure*}[tbp]
\begin{center}
\includegraphics[width=1.0\columnwidth,angle=0]{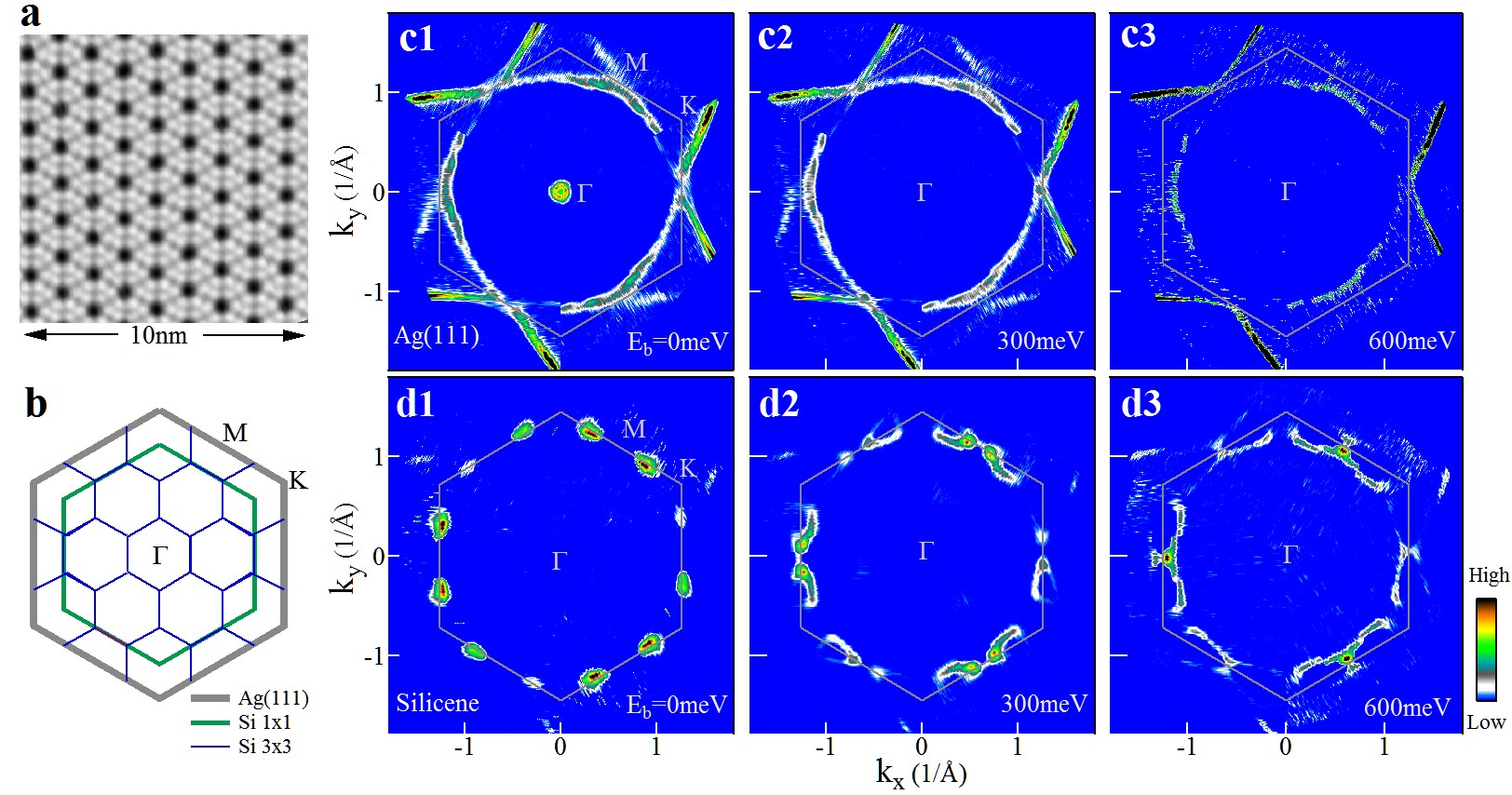}
\end{center}
\caption{Constant energy contours of silicene(3$\times$3)/Ag(111) showing the existence of six pairs of Dirac cones.  (a) STM image of silicene(3$\times$3) grown on the Ag(111) surface. (b) First Brillouin zone of Ag(111) (thick gray solid line),  and the corresponding Brillouin zones of silicene(1$\times$1) (green line) and (3$\times$3) supercell (thin blue line). Here silicene(3$\times$3) is named with respect to the primary silicene(1$\times$1) structure (it is named silicene (4$\times$4) with reference to the Ag(111) surface).  (c1-c3) Constant energy contours of Ag(111) surface measured at 20 K obtained by integrating the photoemission spectral weight over a small energy window ($\pm$10 meV) with respect to the binding energy of 0 (c1), 300 meV (c2) and 600 meV (c3). (d1-d3) Constant energy contours of silicene(3$\times$3)/Ag(111) measured at 20 K at three different binding energies of 0 (d1), 300 meV (d2) and 600 meV (d3). Some residual signal of the Ag(111) surface can be discerned which is relatively stronger in the second Brillouin zone. In (c) and (d), the grey line represents the first Brillouin zone of Ag(111) surface. The images are obtained by symmetrizing the original data assuming three-fold symmetry.}
\end{figure*}

\begin{figure*}[tbp]
\begin{center}
\includegraphics[width=1.0\columnwidth,angle=0]{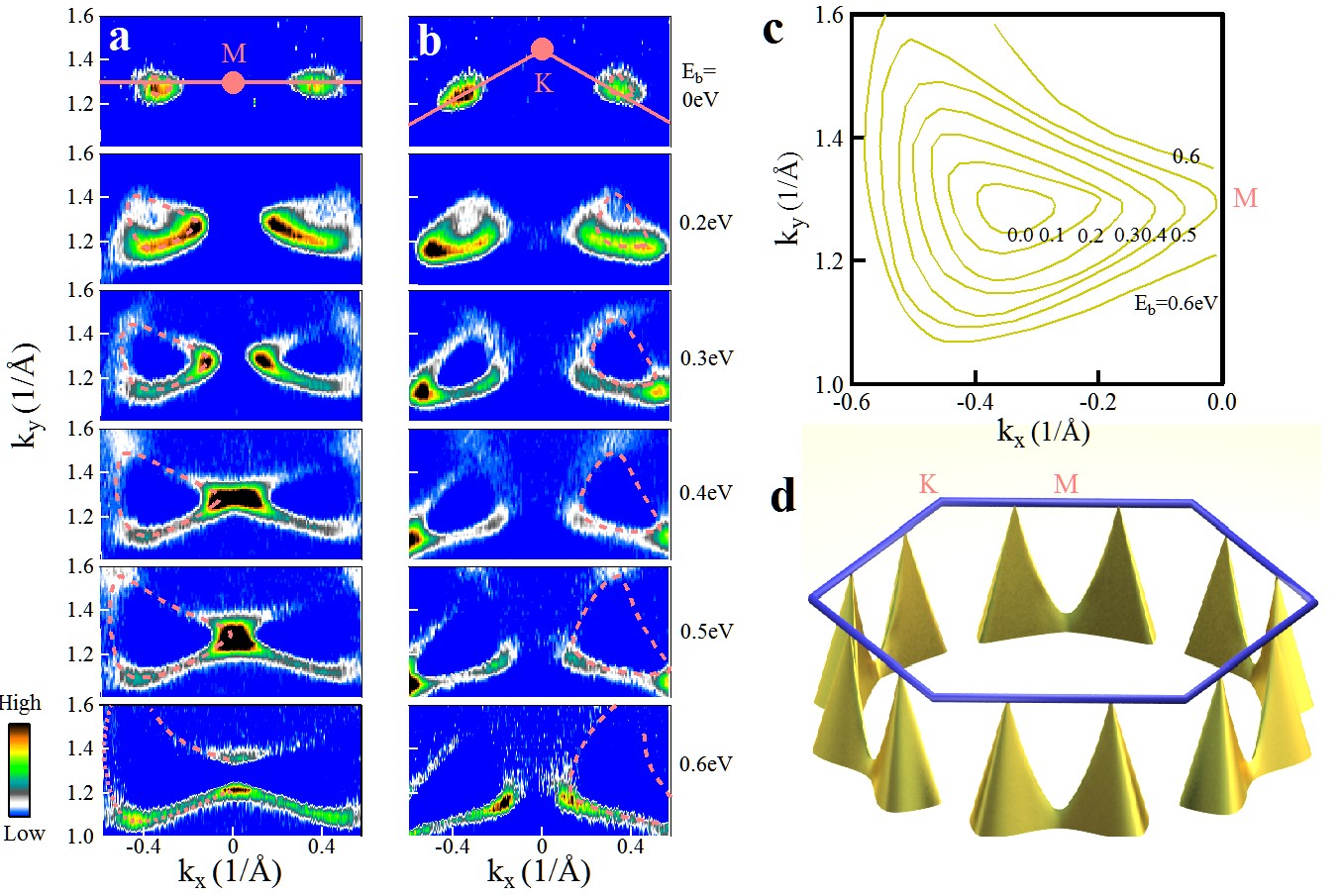}
\end{center}
\caption{Evolution of the Dirac cones in silicene(3$\times$3)/Ag(111) with binding energy.  (a) Constant energy contours of one pair of Dirac cones around the M point obtained by integrating the photoemission spectral weight over a small energy window ($\pm$10 meV) for binding energies of 0, 200 meV, 300 meV, 400 meV, 500meV and 600 meV (from the top to the bottom panels). (b) Constant energy contours of two Dirac cones around the K point at different binding energies. (c) Constant energy contours of a single Dirac cone obtained by the contour lines at different binding energies of 0, 200 meV, 300 meV, 400 meV, 500meV and 600 meV (dashed orange lines in (a) and (b)). (d) Schematic three-dimensional diagram showing the existence of twelve Dirac cones in silicene(3$\times$3)/Ag(111) forming six pairs along the first Brillouin zone edges of the Ag(111) surface (thick blue line).}
\end{figure*}

\begin{figure*}[tbp]
\begin{center}
\includegraphics[width=1.0\columnwidth,angle=0]{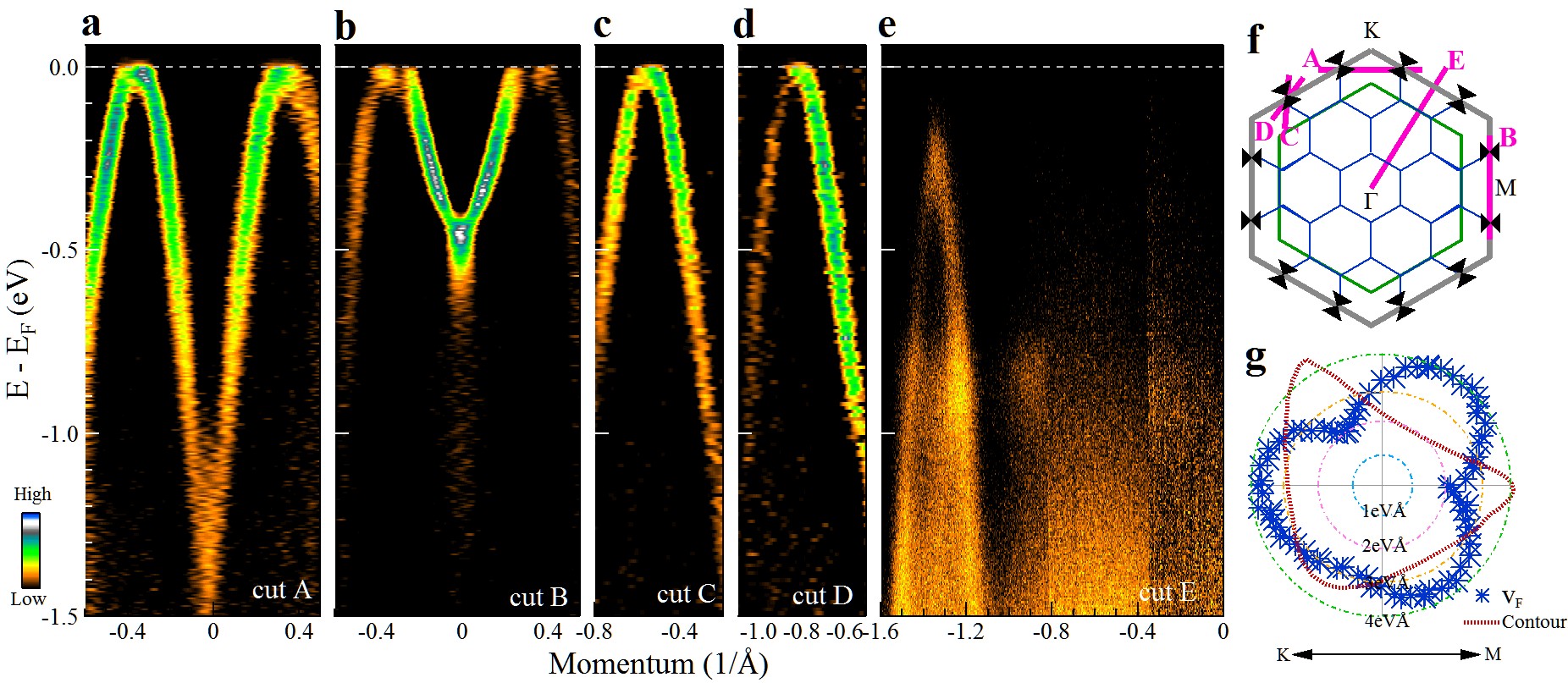}
\end{center}
\caption{Band structures of silicene(3$\times$3)/Ag(111) along different momentum cuts.  (a-e) Band structures measured along five typical momentum cuts. Location of the five momentum cuts is shown in (f). To highlight the measured bands, the images shown are second-derivative images of the original data with respect to the momentum.  (g) Fermi velocity of the Dirac cone along different directions plotted as blue asterisks in a polar coordinate. The triangle-shaped brown line represents the corresponding constant energy contour line of the Dirac cone at a binding energy of 0.5 eV (see Fig. 2c).}
\end{figure*}

\begin{figure*}[tbp]
\begin{center}
\includegraphics[width=1.0\columnwidth,angle=0]{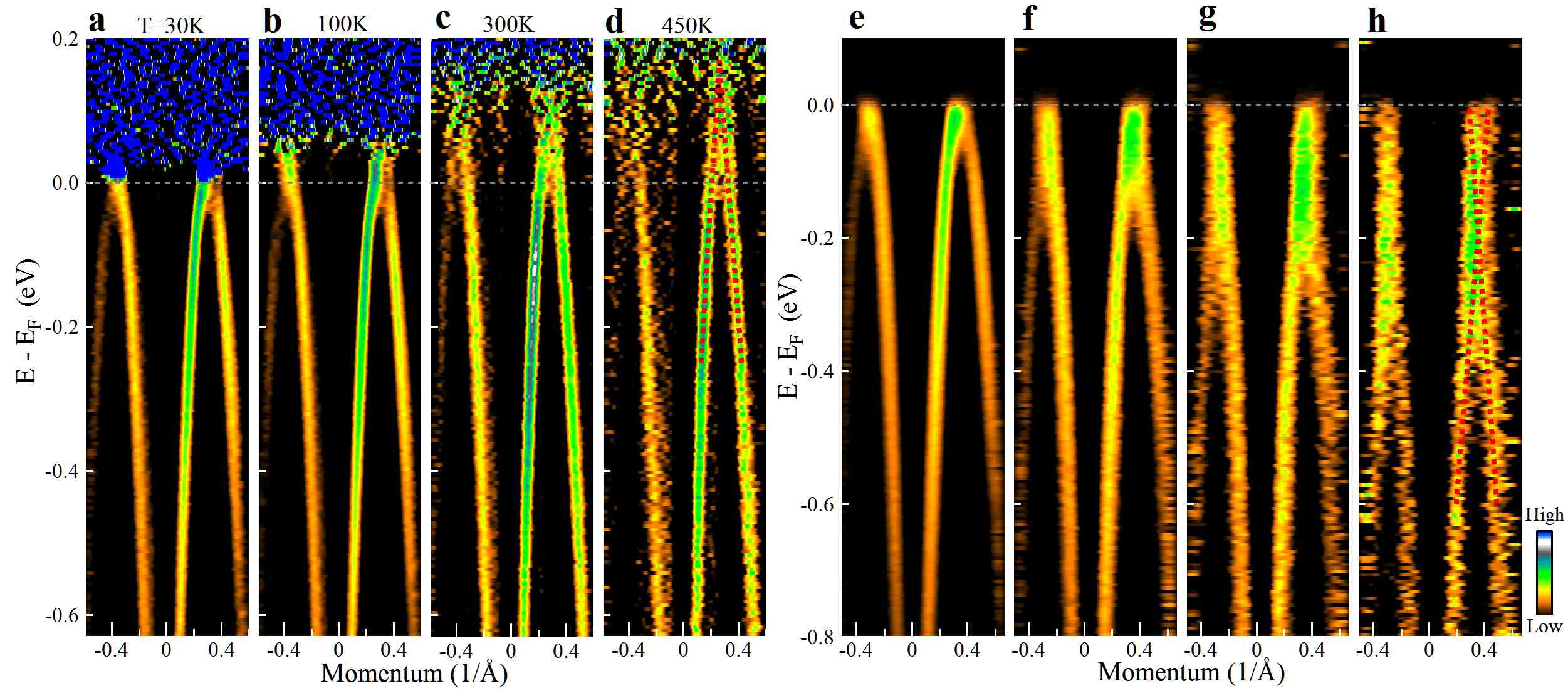}
\end{center}
\caption{Revelation of the Dirac cone and the upper Dirac branch in silicene(3$\times$3)/Ag(111).  (a-d) Band structures measured along the cut A in Fig. 3f at a temperature of 30 K (a), 100 K (b), 300 K (c) and 450 K (d). The images have been divided by the corresponding Fermi-Dirac distribution functions in order to observe band structures above the Fermi level.  (e-h) Band structures measured along the cut A in Fig. 3f after depositing potassium on the surface.  When an increasing amount of potassium is deposited on the sample surface, the overall band structure shifts to higher binding energy because of electron-doping. The red dashed lines in (d) and (h) are guides to the eye through the observed bands.
}

\end{figure*}

\end{document}